\documentclass[aps,prl,twocolumn,superscriptaddress,showpacs,preprintnumbers,amsmath,amssymb]{revtex4}
\usepackage{graphicx} 
\usepackage{dcolumn}  

\graphicspath{{ps}}

\newcommand{\ee}{e^{+}e^{-}}

\newcommand{\pipi}{\pi^{+}\pi^{-}}
\newcommand{\kk}{K^{+}K^{-}}
\newcommand{\piz}{\pi^{0}}
\newcommand{\pip}{\pi^{+}}
\newcommand{\pim}{\pi^{-}}
\newcommand{\bbar}{B\bar{B}}
\newcommand{\qqbar}{q\bar{q}}

\newcommand{\Bz}{B^{0}}
\newcommand{\Bpl}{B^{+}}

\newcommand{\Mbc}{M_{\rm bc}}
\newcommand{\DE}{\Delta E}

\newcommand{\rt}{\rightarrow}

\newcommand{\Dz}{{D}^0}
\newcommand{\Dzbar}{\bar{D}^0}
\newcommand{\Dmi}{D^-}
\newcommand{\Dsz}{D^{*+}_{s0}}
\newcommand{\Ds}{D_{s}^+}
\newcommand{\Zpp}{\mathbf{z^{++}}}

\newcommand{\Zz}{\mathbf{z^{0}}}

\begin{document}



\title{ \quad\\[0.5cm] Measurements of {\boldmath $B\rt \bar{D}\Dsz (2317)$} decay rates and a search for isospin
partners of the {\boldmath $\Dsz (2317)$}}

\noaffiliation
\affiliation{University of the Basque Country UPV/EHU, 48080 Bilbao}
\affiliation{Beihang University, Beijing 100191}
\affiliation{University of Bonn, 53115 Bonn}
\affiliation{Budker Institute of Nuclear Physics SB RAS and Novosibirsk State University, Novosibirsk 630090}
\affiliation{Faculty of Mathematics and Physics, Charles University, 121 16 Prague}
\affiliation{University of Cincinnati, Cincinnati, Ohio 45221}
\affiliation{Deutsches Elektronen--Synchrotron, 22607 Hamburg}
\affiliation{Justus-Liebig-Universit\"at Gie\ss{}en, 35392 Gie\ss{}en}
\affiliation{SOKENDAI (The Graduate University for Advanced Studies), Hayama 240-0193}
\affiliation{Gyeongsang National University, Chinju 660-701}
\affiliation{Hanyang University, Seoul 133-791}
\affiliation{University of Hawaii, Honolulu, Hawaii 96822}
\affiliation{High Energy Accelerator Research Organization (KEK), Tsukuba 305-0801}
\affiliation{IKERBASQUE, Basque Foundation for Science, 48013 Bilbao}
\affiliation{Indian Institute of Technology Guwahati, Assam 781039}
\affiliation{Indian Institute of Technology Madras, Chennai 600036}
\affiliation{Indiana University, Bloomington, Indiana 47408}
\affiliation{Center for Underground Physics, Institute for Basic Science, Daejeon 305-811}
\affiliation{Institute of High Energy Physics, Vienna 1050}
\affiliation{Institute for High Energy Physics, Protvino 142281}
\affiliation{INFN - Sezione di Torino, 10125 Torino}
\affiliation{Institute for Theoretical and Experimental Physics, Moscow 117218}
\affiliation{J. Stefan Institute, 1000 Ljubljana}
\affiliation{Kanagawa University, Yokohama 221-8686}
\affiliation{Institut f\"ur Experimentelle Kernphysik, Karlsruher Institut f\"ur Technologie, 76131 Karlsruhe}
\affiliation{Kennesaw State University, Kennesaw GA 30144}
\affiliation{King Abdulaziz City for Science and Technology, Riyadh 11442}
\affiliation{Korea Institute of Science and Technology Information, Daejeon 305-806}
\affiliation{Korea University, Seoul 136-713}
\affiliation{Kyungpook National University, Daegu 702-701}
\affiliation{\'Ecole Polytechnique F\'ed\'erale de Lausanne (EPFL), Lausanne 1015}
\affiliation{Faculty of Mathematics and Physics, University of Ljubljana, 1000 Ljubljana}
\affiliation{Luther College, Decorah, Iowa 52101}
\affiliation{University of Maribor, 2000 Maribor}
\affiliation{Max-Planck-Institut f\"ur Physik, 80805 M\"unchen}
\affiliation{School of Physics, University of Melbourne, Victoria 3010}
\affiliation{Moscow Physical Engineering Institute, Moscow 115409}
\affiliation{Moscow Institute of Physics and Technology, Moscow Region 141700}
\affiliation{Graduate School of Science, Nagoya University, Nagoya 464-8602}
\affiliation{Kobayashi-Maskawa Institute, Nagoya University, Nagoya 464-8602}
\affiliation{Nara Women's University, Nara 630-8506}
\affiliation{National Central University, Chung-li 32054}
\affiliation{National United University, Miao Li 36003}
\affiliation{Department of Physics, National Taiwan University, Taipei 10617}
\affiliation{H. Niewodniczanski Institute of Nuclear Physics, Krakow 31-342}
\affiliation{Niigata University, Niigata 950-2181}
\affiliation{Osaka City University, Osaka 558-8585}
\affiliation{Pacific Northwest National Laboratory, Richland, Washington 99352}
\affiliation{Peking University, Beijing 100871}
\affiliation{University of Pittsburgh, Pittsburgh, Pennsylvania 15260}
\affiliation{University of Science and Technology of China, Hefei 230026}
\affiliation{Seoul National University, Seoul 151-742}
\affiliation{Soongsil University, Seoul 156-743}
\affiliation{University of South Carolina, Columbia, South Carolina 29208}
\affiliation{Sungkyunkwan University, Suwon 440-746}
\affiliation{School of Physics, University of Sydney, NSW 2006}
\affiliation{Department of Physics, Faculty of Science, University of Tabuk, Tabuk 71451}
\affiliation{Tata Institute of Fundamental Research, Mumbai 400005}
\affiliation{Excellence Cluster Universe, Technische Universit\"at M\"unchen, 85748 Garching}
\affiliation{Toho University, Funabashi 274-8510}
\affiliation{Tohoku University, Sendai 980-8578}
\affiliation{Department of Physics, University of Tokyo, Tokyo 113-0033}
\affiliation{Tokyo Institute of Technology, Tokyo 152-8550}
\affiliation{Tokyo Metropolitan University, Tokyo 192-0397}
\affiliation{University of Torino, 10124 Torino}
\affiliation{CNP, Virginia Polytechnic Institute and State University, Blacksburg, Virginia 24061}
\affiliation{Wayne State University, Detroit, Michigan 48202}
\affiliation{Yamagata University, Yamagata 990-8560}
\affiliation{Yonsei University, Seoul 120-749}
 \author{S.-K.~Choi}\affiliation{Gyeongsang National University, Chinju 660-701} 
 \author{S.~L.~Olsen}\affiliation{Center for Underground Physics, Institute for Basic Science, Daejeon 305-811} 
  \author{A.~Abdesselam}\affiliation{Department of Physics, Faculty of Science, University of Tabuk, Tabuk 71451} 
  \author{I.~Adachi}\affiliation{High Energy Accelerator Research Organization (KEK), Tsukuba 305-0801}\affiliation{SOKENDAI (The Graduate University for Advanced Studies), Hayama 240-0193} 
  \author{H.~Aihara}\affiliation{Department of Physics, University of Tokyo, Tokyo 113-0033} 
  \author{K.~Arinstein}\affiliation{Budker Institute of Nuclear Physics SB RAS and Novosibirsk State University, Novosibirsk 630090} 
  \author{D.~M.~Asner}\affiliation{Pacific Northwest National Laboratory, Richland, Washington 99352} 
  \author{T.~Aushev}\affiliation{Moscow Institute of Physics and Technology, Moscow Region 141700}\affiliation{Institute for Theoretical and Experimental Physics, Moscow 117218} 
  \author{R.~Ayad}\affiliation{Department of Physics, Faculty of Science, University of Tabuk, Tabuk 71451} 
  \author{V.~Babu}\affiliation{Tata Institute of Fundamental Research, Mumbai 400005} 
% ...
\collaboration{The Belle Collaboration}



\begin{abstract}
We report improved measurements of the product branching fractions 
${\mathcal B}(\Bpl\rt\Dzbar\Dsz (2317))\times{\mathcal B}(\Dsz (2317)\rt\Ds\piz)
=(8.0^{+1.3}_{-1.2} \pm 1.1 \pm 0.4)\times 10^{-4}$
and ${\mathcal B}(\Bz\rt\Dmi\Dsz(2317))\times{\mathcal B}(\Dsz (2317)\rt\Ds\piz)
=(10.2^{+1.3}_{-1.2} \pm 1.0 \pm 0.4)\times 10^{-4}$, where the
first errors are statistical, the second are systematic and
the third are from $D$ and $\Ds$ branching fractions.
In addition, we report negative results from a search for
hypothesized neutral ($\Zz$) and doubly charged ($\Zpp$) isospin partners
of the $\Dsz (2317)$ and provide upper limits on the 
product branching fractions
${\mathcal B}(\Bz\rt\Dz \Zz)\times{\mathcal B}(\Zz\rt \Ds \pim)$ and
${\mathcal B}(\Bpl\rt\Dmi \Zpp)\times{\mathcal B}(\Zpp\rt \Ds \pip)$
that are more than an order of magnitude smaller than theoretical expectations
for the hypotheses that the $\Dsz(2317)$ is a member of an isospin triplet. 
The analysis  uses a 711~fb$^{-1}$ data sample containing 772~million
$\bbar$-meson pairs collected at the $\Upsilon(4S)$ resonance in the
Belle detector at the KEKB collider.

\end{abstract}

\pacs{12.39.Mk, 13.20.He, 14.40.Lb}

\maketitle


{\renewcommand{\thefootnote}{\fnsymbol{footnote}}}
\setcounter{footnote}{0}

\section{Introduction}
The $\Dsz(2317)$ meson, hereinafter referred to as the $\Dsz$, was
first observed by BABAR as a narrow peak in the $\Ds\piz$
invariant mass spectrum produced in inclusive  $\ee\rt\Ds\piz X$
annihilation processes~\cite{babar_dsz,conj}, and confirmed
by CLEO~\cite{CLEO_dsz}.  Its production in the $B$-meson
decay processes $B\rt \bar{D}\Dsz$ was subsequently established
by both Belle~\cite{belle_bdsz} and BABAR~\cite{babar_bdsz}.
(Here, $B$ and $\bar{D}$
are used to denote $\Bz$ and $\Dmi$ or  $\Bpl$ and $\Dzbar$.)
Although it is generally considered to be the conventional
$I(J^{P})=0(0^{+})$ $P$-wave $c\bar{s}$ meson, its mass,
$M_{\Dsz}=2317.8\pm 0.6$~MeV~\cite{pdg,c=1}, is the same as the peak mass of its
nonstrange counterpart, the $0^+$ $P$-wave $c\bar{q}$ ($q=u$~or~$d$) 
$D^{*}_0$ with mass $M_{D^{*}_0}= 2318 \pm 29$~MeV~\cite{pdg}, in spite of the
fact that the mass of the $s$ quark is $\sim 100$~MeV above that of either
of the $q$ quarks.  
Potential-model~\cite{potential_models} and lattice QCD~\cite{lqcd}
calculations published prior to the BABAR discovery predicted that the $0^+$ $P$-wave
$c\bar{s}$ meson mass would be well above the  $m_{D^0} + m_{K^+} = 2358.6$~MeV
threshold and have a large partial decay width for the strong-interaction-
allowed process $\Dsz\rt DK$.  The observation of a subthreshold mass has
led to theoretical speculation that the $\Dsz$ is not a simple $c\bar{s}$ meson,
but instead a $DK$ molecule~\cite{molecule},
a diquark-diantiquark state~\cite{diquark} or some mixture of a $c\bar{s}$ core state with a
$DK$ molecule and/or a diquark-diantiquark~\cite{mixed_molecule}.  

A $c\bar{s}$ meson with mass below the
$2358.6$~MeV threshold would decay via the isospin-violating process 
$\Dsz\rt\Ds\piz$ or the electromagnetic process $\Dsz\rt D^{*+}_s \gamma$ and, thus,
have a narrow natural width.  This is consistent with experimental measurements, which have
established a 95\% C.L. upper limit on the total width of $\Gamma_{\Dsz}\le 3.8$~MeV~\cite{pdg}.
The small width of the $\Dsz$ is evidence for an $I=0$ assignment.  However, the CLEO
experiment has established a stringent 90\% C.L. upper limit on the partial width for
$\Dsz\rt \Ds \gamma$ decay~\cite{CLEO_dsz}:
\begin{equation}
R(\Dsz ) \equiv \frac{\Gamma(\Dsz\rt \Ds \gamma) }{\Gamma(\Dsz\rt\Ds \piz)}\le 0.059,
\end{equation}
while studies that consider the $\Dsz$ to be the $c\bar{s}$ chiral partner of the 
$\Ds$~\cite{mehen} predict values for $R(\Dsz)$ that are higher than the CLEO
upper limit. Product branching fractions for $B\rt \bar{D}\Dsz ,~~\Dsz\rt\Ds\piz$
have been measured by BABAR~\cite{babar_bdsz} and Belle~\cite{belle_bdsz};
the Particle Data Group (PDG) averages~\cite{pdg} of their results are:
\begin{eqnarray}
{\mathcal B}(\Bpl\rt\Dzbar\Dsz)\times{\mathcal B}(\Dsz\rt\Ds\piz) &=& (7.3^{+2.2}_{-1.7})\times 10^{-4}, 
\nonumber \\
{\mathcal B}(\Bz\rt\Dmi\Dsz)\times{\mathcal B}(\Dsz\rt\Ds\piz) &=& (9.7^{+4.0}_{-3.3})\times 10^{-4}.
\nonumber
\end{eqnarray}
Under the plausible assumption that ${\mathcal B}(\Dsz\rt\Ds\piz)\sim 1$,
these measurements translate into the branching fraction ratios
\begin{eqnarray}
\frac{{\mathcal B}(\Bpl\rt\Dzbar\Dsz)}{{\mathcal B}(\Bpl\rt \Dzbar\Ds)} &=& 0.081^{+0.026}_{-0.021}, 
\nonumber \\
\frac{{\mathcal B}(\Bz\rt\Dmi\Dsz)}{{\mathcal B}(\Bz\rt \Dmi\Ds)} &=&  0.13^{+0.06}_{-0.05},
\nonumber
\end{eqnarray}
which the authors of Refs.~\cite{datta} and~\cite{hsiao-nan} note are well below expectations for a
purely $c\bar{s}$ quark-antiquark state and an indication of some kind of multiquark content. 

The BABAR and Belle measurements for both $\Bpl$ and $\Bz$ modes agree within errors, 
the biggest difference is $1.5\sigma$ for the $\Bz$ mode.
In both cases, the measurements are based
on event samples that are about 20\% of the currently available data.  Updated measurements
based on the full data sets from both experiments would be useful.

A report by Hayashigaki and Terasaki~\cite{hayashigaki} concluded that an 
$I=1$ and $I_3=0$ assignment for the $\Dsz$ cannot be ruled out and claimed, in fact, 
that an $I=1$ diquark-diantiquark interpretation is favored by some existing data.  
If this were the case, doubly charged $I_3=1$ ($\Zpp)$ and neutral $I_3=-1$ 
($\Zz$) partners of the $\Dsz$ with mass within $\sim\pm 10$~MeV of $M_{\Dsz}$ should exist.
Since the $\Zpp$ and $\Zz$ would be charmed mesons with $I=1$ and $S=1$,
they would necessarily have a minimal quark content of $c\bar{s} u\bar{d}$ and
$c\bar{s}d\bar{u}$, respectively.  Although a BABAR search for doubly charged and
neutral partners of the $\Dsz$ in inclusive $\ee$ annihilation events sets $95\%$
C.L. upper limits on their production rates at $1.7\%$ and $1.3\%$, respectively, 
of that for the $\Dsz$~\cite{babar_fplpl}, Terasaki has argued that these do
not conclusively rule out their existence~\cite{terasaki05}.  
If the $\Zpp$ and $\Zz$ mesons exist, isospin invariance ensures that the
product branching fractions  
${\mathcal B}(B\rt \bar{D} \mathbf{z^{++,0}})\times{\mathcal B}(\mathbf{z^{++,0}}\rt\Ds\pi^{+,-})$ 
will be nearly equal to
${\mathcal B}(B\rt \bar{D}\Dsz )\times{\mathcal B}(\Dsz\rt\Ds\piz)$. 

Here, we report measurements of
${\mathcal B}(\Bpl\rt\Dzbar\Dsz)\times{\mathcal B}(\Dsz\rt\Ds\piz)$
and ${\mathcal B}(\Bz\rt\Dmi\Dsz)\times{\mathcal B}(\Dsz\rt\Ds\piz)$
using a data sample that is more than 6 times larger than that used in previous results~\cite{belle_bdsz}
and a search for doubly charged ($\Zpp$)  and neutral ($\Zz$) isospin partners
of the $\Dsz$ in the decay processes
 $\Bpl\rt \Dmi \Zpp ,~~\Zpp\rt\Ds\pip $ 
and $\Bz\rt \Dzbar \Zz ,~~\Zz\rt\Ds\pim $.
The results are based on the full  Belle $\Upsilon(4S)$ data 
sample (711~fb$^{-1}$) that contains 772 million $\bbar$-meson
pairs produced at a center-of-mass system (cms) energy of $\sqrt{s}=10.58$~GeV and collected 
in the Belle detector at the KEKB energy-asymmetric  $\ee$ collider~\cite{KEKB}.  

\section{Detector description}
The Belle detector is a large-solid-angle magnetic 
spectrometer  that consists of a silicon vertex 
detector, a 50-layer cylindrical drift chamber, an 
array of aerogel threshold Cherenkov counters,  a 
barrel-like arrangement of time-of-flight  scintillation 
counters, and an electromagnetic calorimeter
comprised of CsI(Tl) crystals  located inside
a superconducting solenoid coil that provides a 1.5~T
magnetic field.  An iron flux-return located outside of 
the coil is instrumented to detect $K_L$ mesons and to 
identify muons.  The detector is described in detail 
elsewhere~\cite{Belle}.

\section{Event selection}

We reconstruct $\Ds$ mesons via their $\pip\kk$ decay mode, which has
a branching fraction of ${\mathcal B}_{\Ds}=(5.39 \pm 0.21)\%$, $\Dmi$
mesons via the $K^+\pim\pim$ decay mode [${\mathcal B}_{\Dmi}=(9.13\pm0.19)\%$]
and $\Dzbar$ mesons via the $K^+\pim$ [${\mathcal B}_{K\pi}=(3.88\pm0.05)\%$] and
$K^+\pipi\pim$ [${\mathcal B}_{K3\pi}=(8.08\pm 0.20)\%$] decay modes~\cite{pdg}.

For all charged particles, we require $dr<0.7$~cm and $|dz|<3.0$~cm, 
where $d r$ and $d z$ are the track's distances of closest approach
to the run-dependent mean interaction point transverse to and parallel
to the $e^+$ beam direction, respectively.  Charged-particle identification
is accomplished by combining information from different detector
subsystems to form likelihood ratios, 
$L_{K/\pi}=L_{K}/(L_{K}+L_{\pi})$, 
where $L_{K}~(L_{\pi})$ is the likelihood of the kaon (pion)~\cite{Belle-PID}. 
A charged track is classified as a kaon
(pion) if $L_{K/\pi(\pi/K)} >0.5$, with both the muon likelihood ratio 
and electron likelihood smaller than 0.95. 
For $\Bz\rt\Dmi\Dsz$ decay, the kaon and pion identification efficiencies both exceed 95\%.
We reconstruct $\piz$ mesons via their $\piz\rt\gamma\gamma$ decay mode
using $\gamma$ candidates with $E_{\gamma}>30$~MeV and $\gamma\gamma$ 
combinations that satisfy a one-constraint (1C) kinematic fit to 
$m_{\piz}$ with $\chi^2<6.0$.  In addition, we require
$|M_{\gamma\gamma}-m_{\piz}|<12$~MeV and the $\piz$ three-momentum
in the $\ee$ cms $p^{\rm cms}_{\piz}<1.9$~GeV.

Candidate $\bar{D}$ mesons are required to have a $K n\pi$ ($n=1~{\rm to}~3$)
invariant mass in the range $|M_{Kn\pi} -m_{D}|<2.5\sigma$
of the observed peak mass, where $\sigma$ is the width from
a Gaussian fit to the $Kn\pi$ invariant mass peak; $\Ds$
candidates are required to be in the mass interval 
$|M_{\kk\pip}-m_{\Ds}|<2.5\sigma$. 
Here, the values of $\sigma$ range from 4.6~MeV to 5.5~MeV.

Candidate $B \rt \bar{D} \Dsz $ decays are identified by {\it i)} the 
cms energy difference  $\DE\equiv E_B^{\rm cms} - E_{\rm beam}^{\rm cms}$,
{\it ii)} the beam-energy constrained mass 
$\Mbc\equiv\sqrt{(E_{\rm beam}^{\rm cms})^2-(p_B^{\rm cms})^2}$, and
{\it iii)} the $\Ds\piz$ invariant mass.
Here $E_{\rm beam}^{\rm cms}$ is the cms beam energy and $E_B^{\rm cms}$
and $p_B^{\rm cms}$ are the total cms energy and three-momentum of the
particles forming the $\bar{D} \Dsz $ combination.  We select 
events with $M_{\rm bc}>5.20$~GeV, -$0.12~{\rm GeV}<\DE<0.1$~GeV
and $2.228~{\rm GeV}<M_{\Ds\piz}<2.418~{\rm GeV}$ for three-dimensional
fitting, and  define signal regions 
as  $|\Mbc -m_{B}| <0.007$~GeV, $ -0.033~{\rm GeV} < \DE < 0.030$~GeV 
and $|M_{\Ds\piz} - 2.3178~{\rm GeV}|<0.015$~GeV.
For candidate $B\rt \bar{D} \Zpp \ (\Zz)$ decays, the $\piz$ is replaced by a
$\pip \ (\pim)$ and the $ \DE $ signal region is compressed to   
$ | \DE | < 0.023$~GeV.
These intervals correspond approximately to $ \pm 2.5\sigma$ windows around 
the central values for each variable.

To reduce background from $\ee\rt\qqbar$ continuum processes,
where $q=u, d, s, c$, we require the following: $R_2<0.3$, where $R_2$
is the normalized second Fox-Wolfram moment~\cite{fox}; 
$|\cos\theta_B|<0.8$, where $\theta_B$
is the polar angle of the candidate $B$-meson direction
in the cms;  and $|\cos\theta_{{\rm thr}B}|<0.8$,
where $\theta_{{\rm thr}B}$ is the cms angle between the 
thrust axis of the $B$ candidate and that of the remaining unused tracks in the event.
These requirements reject 14\% of $\Bz\rt\Dmi\Dsz$ signal
and 45\% of $\qqbar$ continuum. 

\section{MC simulation}

We use Monte Carlo (MC) simulation to optimize selection criteria, determine
acceptance and study multiple candidates per event~\cite{MC}.  We generate signal
MC for each process under investigation using PDG values~\cite{pdg} for
subdecay branching fractions and setting 
${\mathcal B}(\Dsz\rt \Ds\piz)$ and
${\mathcal B}(\mathbf{z^{++,0}}\rt \Ds\pi^{+,-})=1$.  In addition, we
use a generic $\bbar$ MC sample with about 3 times the 
integrated luminosity of the actual data sample to investigate
possible peaking backgrounds. The simulated events are
processed through the same reconstruction and selection
codes that are used for the real data.

\section{Multiple candidates}

The $\Dsz\rt\Ds\piz$ mode is plagued by a large fraction of events
with multiple candidates.  
The numbers of events with multiple entries in the full fitted region are 
summarized in Table~\ref{tbl:mult}.
Since the MC samples reproduce the
data reasonably well, we use the MC as a guide for methods to
reduce the multiple candidates.

\begin{table}[htb]
\begin{center}
\caption{\label{tbl:mult}
Fractions of multiple candidate events in data and MC.  }
\begin{tabular}{l|c|c|c}\hline  
Sample               &$\Bz\rt\Dmi\Dsz$         & $\Bpl\rt\Dzbar\Dsz$   &  $\Bpl\rt\Dzbar\Dsz$        \\
                     & $\Dmi\rt K\pi\pi$        &  $\Dzbar\rt K\pi$   & $\Dzbar\rt K 3\pi$    \\
\hline\hline 
Sig. MC              &  70\%                   &         45\%           &   70\%                     \\
$\bbar$ MC           &  69\%                   &         39\%           &   69\%                     \\
Data                 &  68\%                   &         39\%           &   69\%                     \\
\hline 

\hline\hline
\end{tabular}
\end{center}
\end{table}

For the $\Dmi\rt K^+\pim\pim$ and $\Dzbar\rt K^+\pipi\pim$ modes,
about two thirds of the multiple candidates are low-energy
photons forming multiple $\piz\rt\gamma\gamma$
combinations and one third are multiple charged pions in the $D$ candidate.
For the $\Dzbar\rt K^+\pim$ mode, essentially all of the multiple candidates
are associated with the $\piz\rt\gamma\gamma$ reconstruction.

\begin{figure}[htb]
\mbox{
  \includegraphics[height=0.15\textwidth,width=0.22\textwidth]{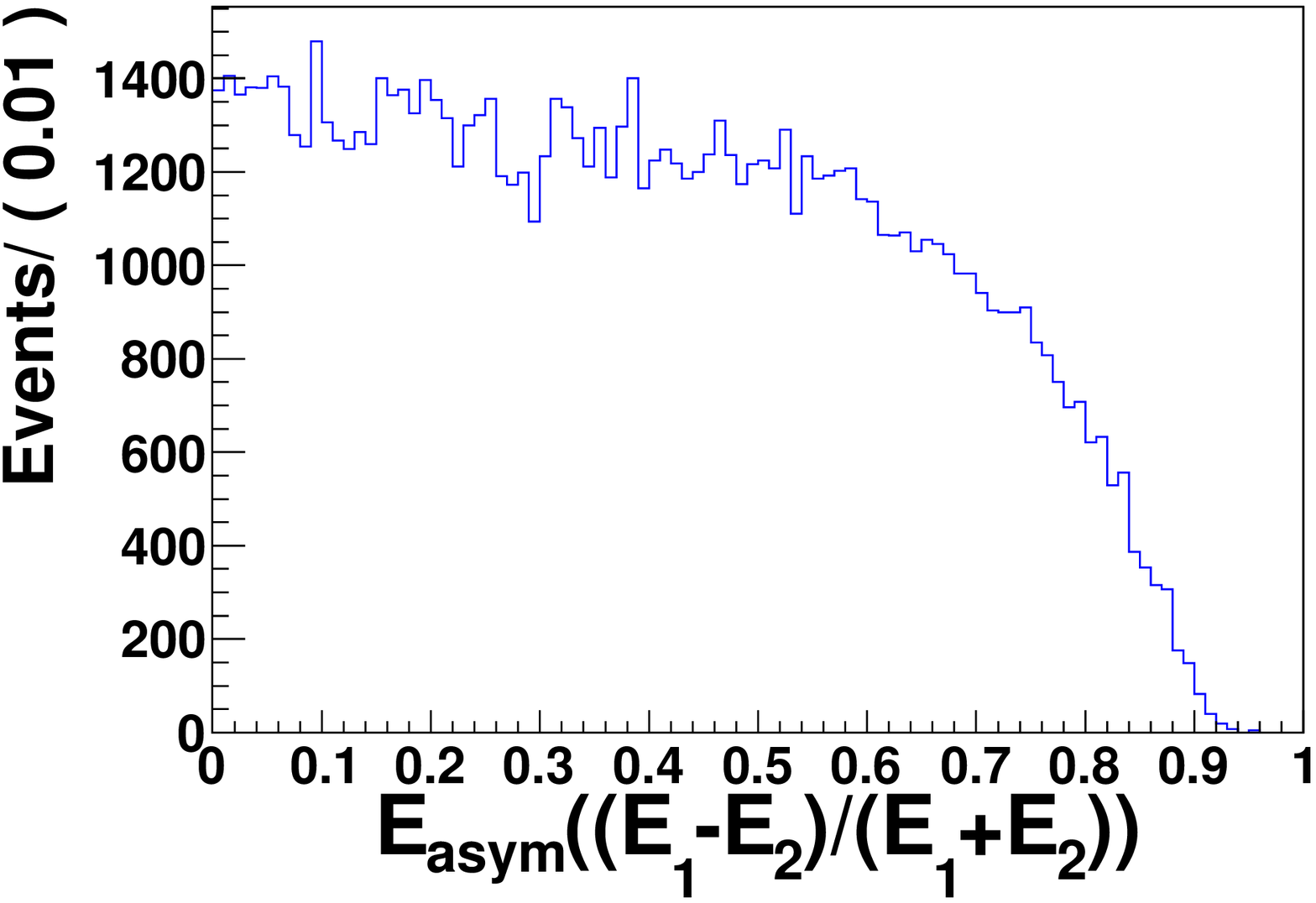}
}
\mbox{
  \includegraphics[height=0.15\textwidth,width=0.22\textwidth]{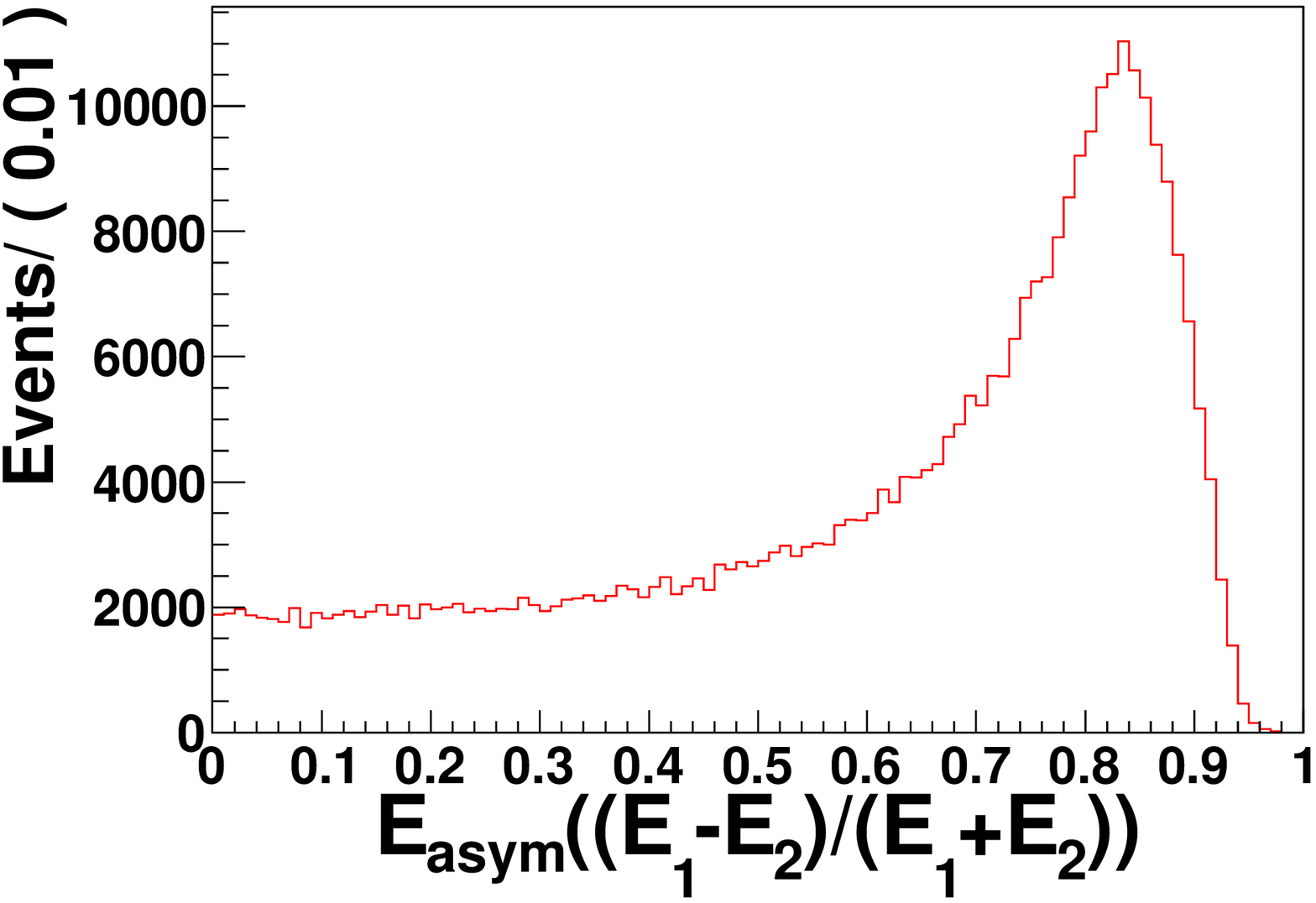}
}
\caption{ 
The $E_{\rm asym}$ distributions for signal MC events for correctly {\bf (left)}
and incorrectly {\bf (right)} assigned photons.
}
\label{fig:easym}
\end{figure}

\begin{figure}[htb]
\mbox{
  \includegraphics[height=0.15\textwidth,width=0.22\textwidth]{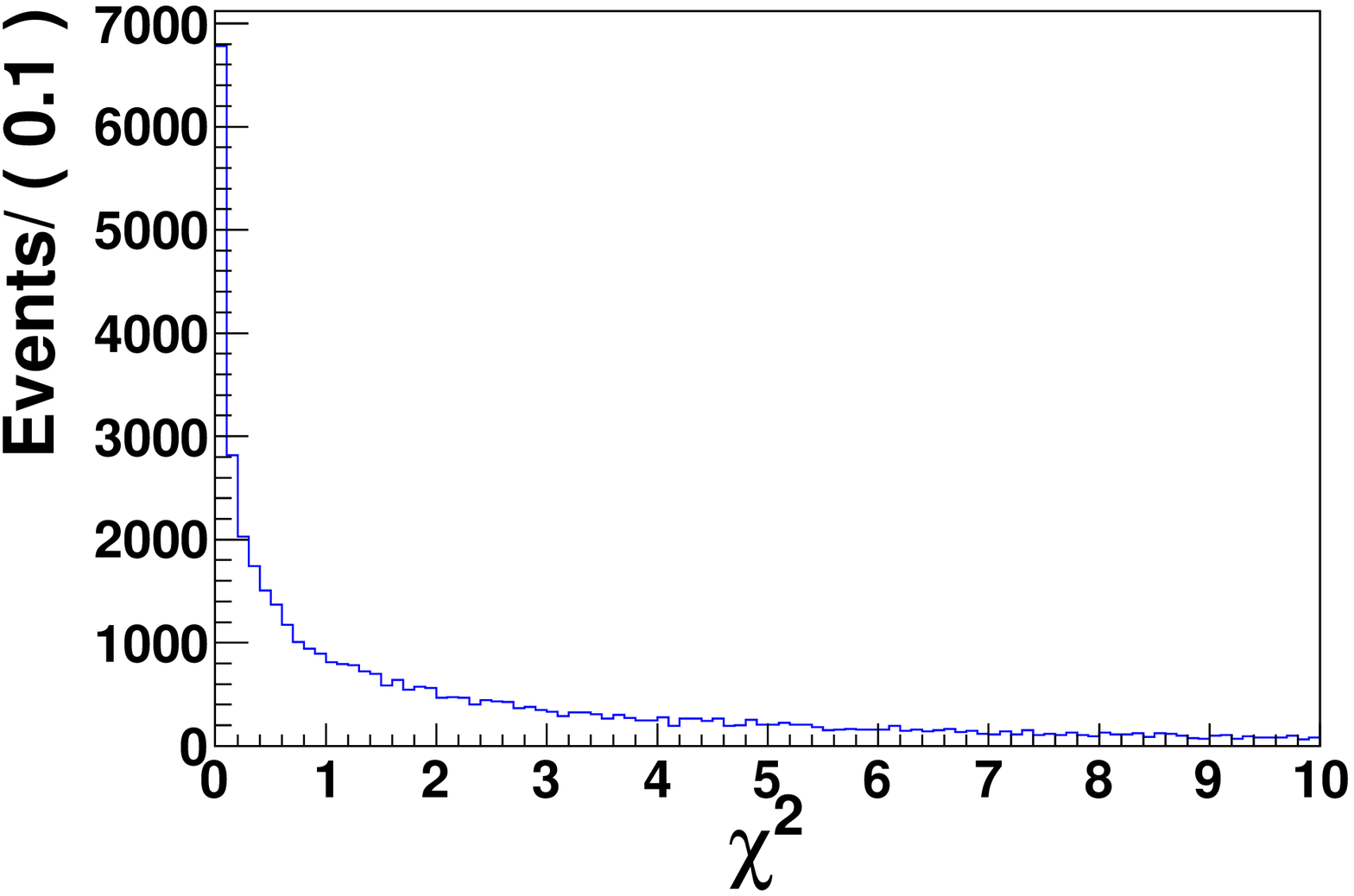}
}
\mbox{
  \includegraphics[height=0.15\textwidth,width=0.22\textwidth]{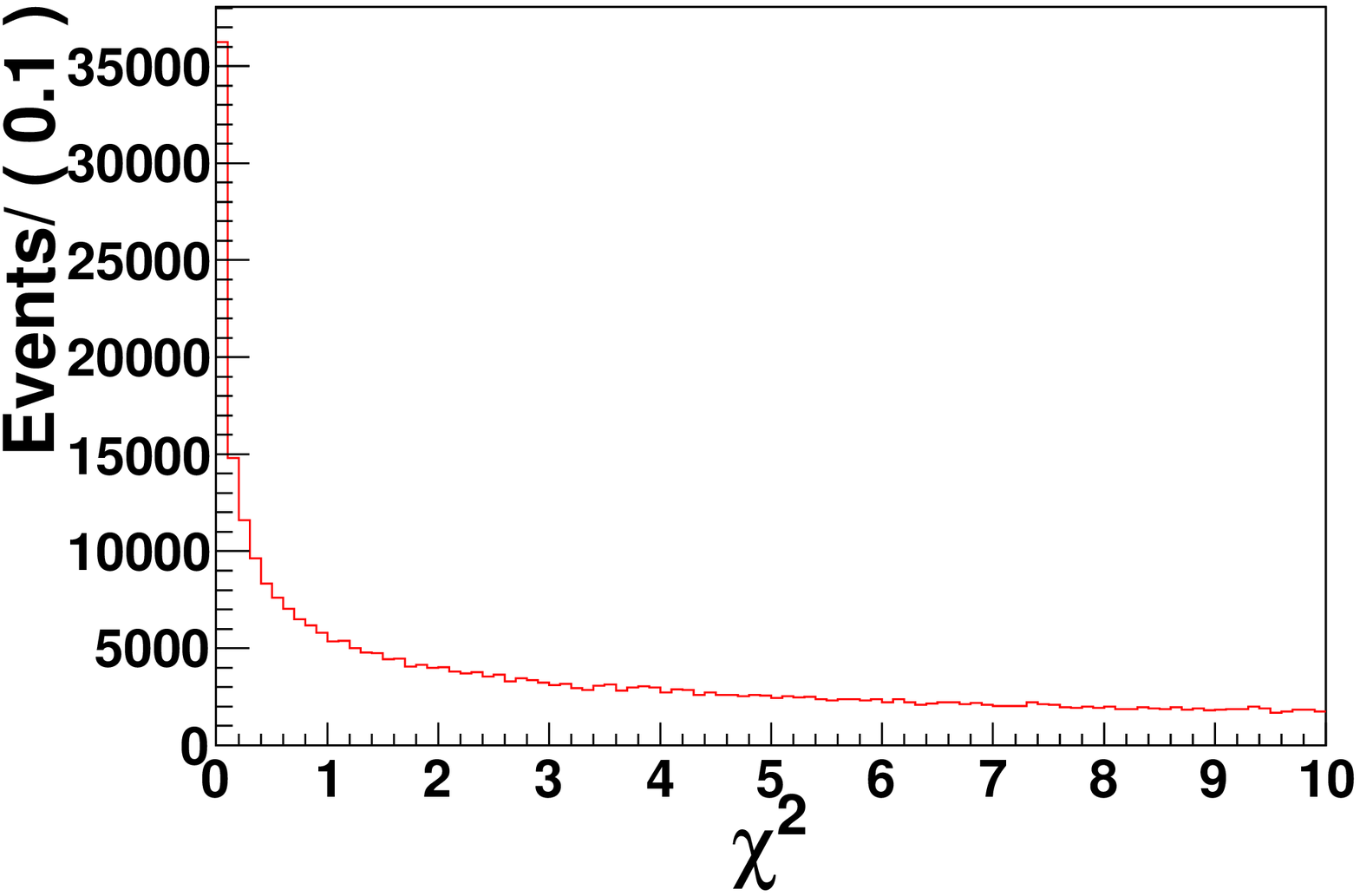}
}
\caption{ 
The $\chi ^2$ distributions from the $\piz\rt\gamma\gamma$ fit for signal MC events
for correctly {\bf (left)} and incorrectly {\bf (right)} assigned photons.
}
\label{fig:chisq}
\end{figure}

We use the $\gamma\gamma$ energy asymmetry, 
$E_{\rm asym}\equiv (E_{1}-E_{2})/(E_{1}+E_{2})$, where $E_{1}$ ($E_{2}$) is the
higher-(lower-)energy photon of the $\gamma\gamma$ pair, to select $\piz$ candidates.
The left panel of Fig.~\ref{fig:easym}
shows the $E_{\rm asym}$ distribution for correctly assigned
$\gamma\gamma$ pairs in signal MC events; the right panel in the same figure shows the
same distribution for incorrectly assigned combinations. Here, the events are required to be in
the $\Mbc$ and $\DE$ signal regions.  According to MC studies, the strong
peak near $E_{\rm asym}\simeq 0.85 $ in the incorrect-assignment plot is mostly due to 
beam-produced background photons.  
Figure~\ref{fig:chisq} shows the corresponding $\chi ^2$ distributions from
the $\piz\rt\gamma\gamma$ kinematic fits.
To reduce the $\gamma$-associated multiple candidates while
minimizing loss of signal efficiency, we require that photons in the energy interval
$30~{\rm MeV}<E_{\gamma}<40~{\rm MeV}$
have $\chi^2<0.5$ for the 1C fit or $E_{\rm asym}<0.7$.
For remaining events with multiple $\gamma$ candidates, we select the combination with the
smallest $E_{\rm asym}$ value.   For multiple $\bar{D}$ ($\Ds$) candidates, we select
the track combination with invariant mass closest to the PDG value for $m_D$ ($m_{\Ds}$). 

\section{{\boldmath $\bar{D}D^{*+}_{{\small s} 0}$} efficiencies}  

We determine event yields from unbinned three-dimensional likelihood fits
[$\Mbc$ {\it vs } $M(\Ds\piz)$ {\it vs } $\DE$] to the selected data using
a bifurcated Gaussian function for the $\Mbc$ signal probability density
function (PDF) and an ARGUS function~\cite{ARGUS} multiplied by a 
second-order Chebyshev polynomial for the $\Mbc$ combinatorial-background PDF.  For
$\DE$, we use a Crystal Ball function~\cite{Xtal-ball} for the signal PDF
and a third-order Chebyshev polynomial for the combinatorial-background PDF.
For $M(\Ds\piz)$, we use a Gaussian 
function for the signal PDF and a third-order
Chebyshev polynomial for the combinatorial-background PDF.

In the generic $\bbar$ MC samples, there is background that peaks in $\Mbc$
and $\DE$ [but not $M(\Ds\piz)$] mostly coming from three-body $B\rt \bar{D}\piz\Ds$
decays.  This background is modeled in the fits by $\Mbc$ and $\DE$ signal
functions and a linear function for $M(\Ds\piz)$.

\begin{figure}[htb]
\mbox{
  \includegraphics[height=0.24\textwidth,width=0.47\textwidth]{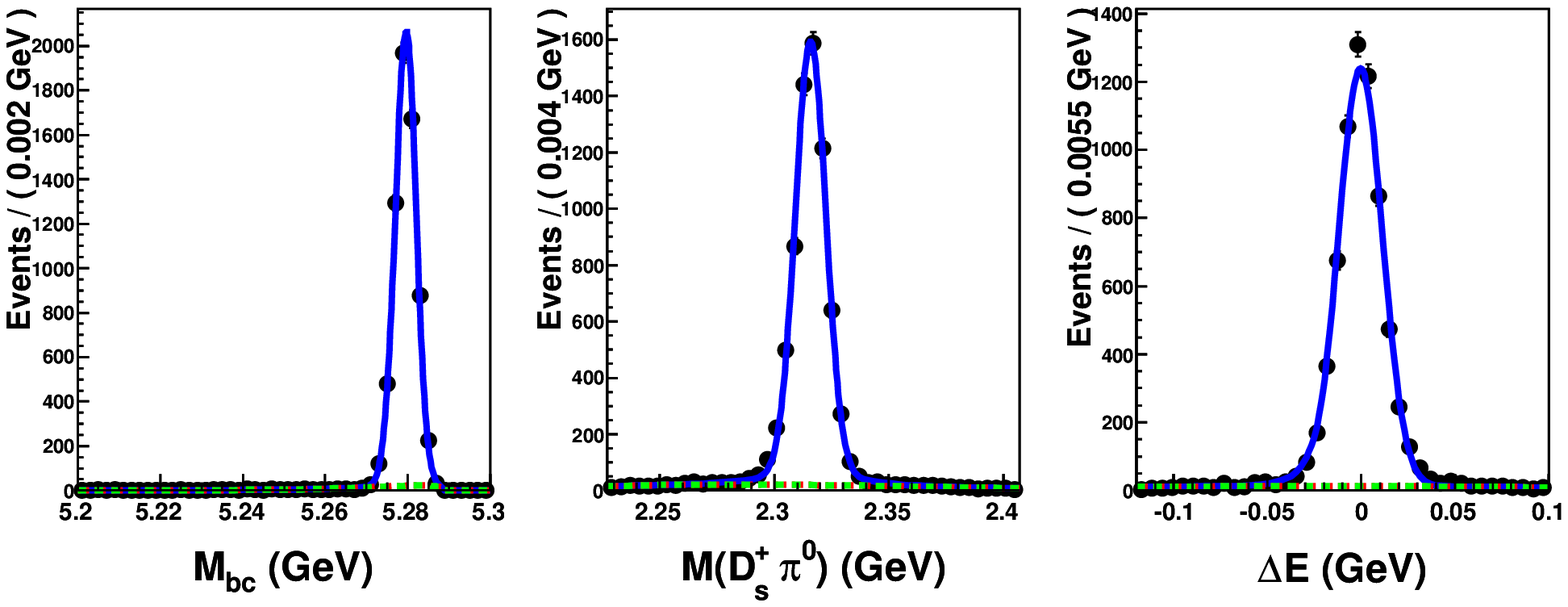}
}
\mbox{
  \includegraphics[height=0.24\textwidth,width=0.47\textwidth]{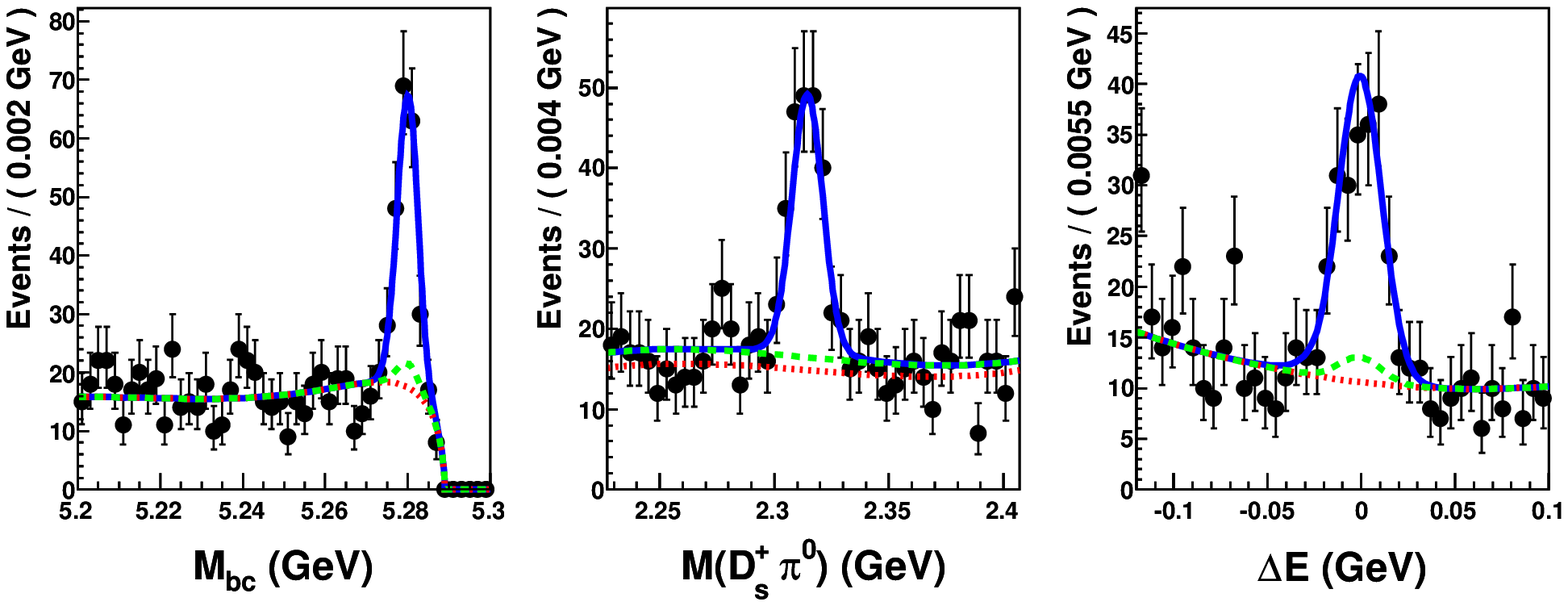}
}
\caption{ {\bf Top:}
The $\Mbc$ (left), $M(\Ds\piz)$ (center) and $\DE$ (right) distributions
for the $\Bz\rt\Dmi\Dsz$ signal MC events with the results of the fit
superimposed.  The events in each distribution are
in the signal regions of the two quantities not being plotted.
{\bf Bottom:} The corresponding distributions for the generic MC event sample
($\sim$ 3 times the data).
(See text for curves.) }
\label{fig:MCfits}
\end{figure}

As an example, we show fit results for the $\Bz\rt\Dmi\Dsz$ signal MC sample in
the upper part of Fig.~\ref{fig:MCfits}.  
The lower part of Fig.~\ref{fig:MCfits} shows the results
from fits to the generic MC sample.  In these figures and subsequent plots
in this report, the red short-dashed curve is the
fitted background; the green long-dashed curve has the peaking background
added and the solid blue curve includes the signal.

The detection efficiencies determined from the signal MC events that survive the application
of the multiple event selection requirements are listed in Table~\ref{tbl:piz_effic}.

\begin{table}[htb]
\begin{center}
\caption{\label{tbl:piz_effic}
The MC-determined $B\rt\bar{D}\Dsz$ efficiencies.}  
\begin{tabular}{l|c|c|c}\hline  
                               &$\Bz\rt\Dmi\Dsz$         & $\Bpl\rt\Dzbar\Dsz$   &  $\Bpl\rt\Dzbar\Dsz$        \\
                               & $\Dmi\rt K^+\pim\pim$   &  $\Dzbar\rt K^+\pim$   & $\Dzbar\rt K^+\pipi\pim$    \\
\hline\hline 
$N_{\rm gen}$     &  266230                 &         266230        &   266230                   \\
$N_{\rm fit}$     &  $7022\pm 90$           &     $8575\pm 97$      &   $4839\pm72$              \\
effic.          &  $(2.64\pm 0.03)$\%     &  $(3.22\pm 0.04)$\%   &   $(1.82\pm0.03)$\%        \\
\hline 

\hline\hline
\end{tabular}
\end{center}
\end{table}

\section{{\boldmath $B\rt \bar{D}\Dsz$};~~ {\boldmath $\Dsz\rt\Ds\piz$} results}

\subsection{1)  {\boldmath $\Bz\rt\Dmi\Dsz$},~~{\boldmath $\Dsz\rt\Ds\piz$}}

\begin{figure}[htb]
\mbox{
  \includegraphics[height=0.24\textwidth,width=0.47\textwidth]{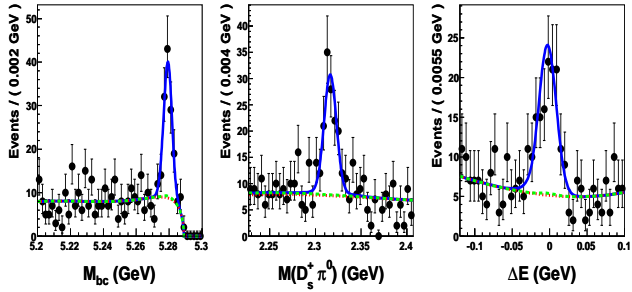}
}
\caption{
The $\Mbc$ (left), $M(\Ds\piz)$ (center) and $\DE$ (right) distributions
for projections of the $\Bz\rt\Dmi\Dsz$ candidate events
that are in the signal regions of the two quantities not being plotted.
The results of the fit described in the text are superimposed.  
(See text for curves.) }

\label{fig:b0-data-fits}
\end{figure}

We determine the number of $\Bz\rt\Dmi\Dsz$;~$\Dsz\rt\Ds\piz$ 
signal events in the data by applying the three-dimensional fit
described above to the selected $\bar{D}=\Dmi$ candidates.  In this
fit, the rms widths of the $\Mbc$, $M(\Ds\piz)$ and $\DE$ signal
functions are kept fixed at their MC-determined values.
Figure~\ref{fig:b0-data-fits} shows the
results of the fit,
which returns a signal yield of $N_{\rm evt}=102.6^{+12.7}_{-12.0}$ events.
The fitted peaking background yield is consistent with zero: $7.7\pm 13.6$
events.  The signal significance, determined as the 
square root of twice the difference of log-likelihood values from fits with and without a signal
term, is $9.9\sigma$.

We determine the product branching fraction from the relation
\begin{eqnarray}
&&{\mathcal B}(\Bz\rt\Dmi\Dsz)\times{\mathcal B}(\Dsz\rt\Ds\piz)
\label{eqn:bf} \\
&=&\frac{N_{\rm evt}}{N_{\bbar}\eta_{\Dmi \Ds}{\mathcal B}_{\Dmi}{\mathcal B}_{\Ds}},
\nonumber
\end{eqnarray}
where $N_{\bbar} = (772\pm 11)\times 10^{6}$ is the number of $\bbar$ events
in the data sample and $\eta_{{\Dmi}{\Ds}}$ is the MC-determined detection efficiency
for this channel (see Table~\ref{tbl:piz_effic}).  The result is
\begin{eqnarray}
&&{\mathcal B}(\Bz\rt\Dmi\Dsz)\times{\mathcal B}(\Dsz\rt\Ds\piz)\\
&=&(10.2^{+1.3}_{-1.2} \pm 1.0 \pm 0.4)\times 10^{-4}, 
\nonumber
\end{eqnarray}
where (and elsewhere in this report) the first error is statistical, the second
is the systematic error (discussed below), and the third reflects the errors on
the PDG branching fractions of the $\Dmi$ and $\Ds$ mesons~\cite{pdg}. 
This result agrees well with the average of the BABAR and previous Belle
measurements mentioned above with a substantial improvement in precision.

\subsection{2)  {\boldmath $\Bpl\rt\Dzbar\Dsz$},~~{\boldmath $\Dsz\rt\Ds\piz$} }

\begin{figure}[htb]
\mbox{
  \includegraphics[height=0.24\textwidth,width=0.47\textwidth]{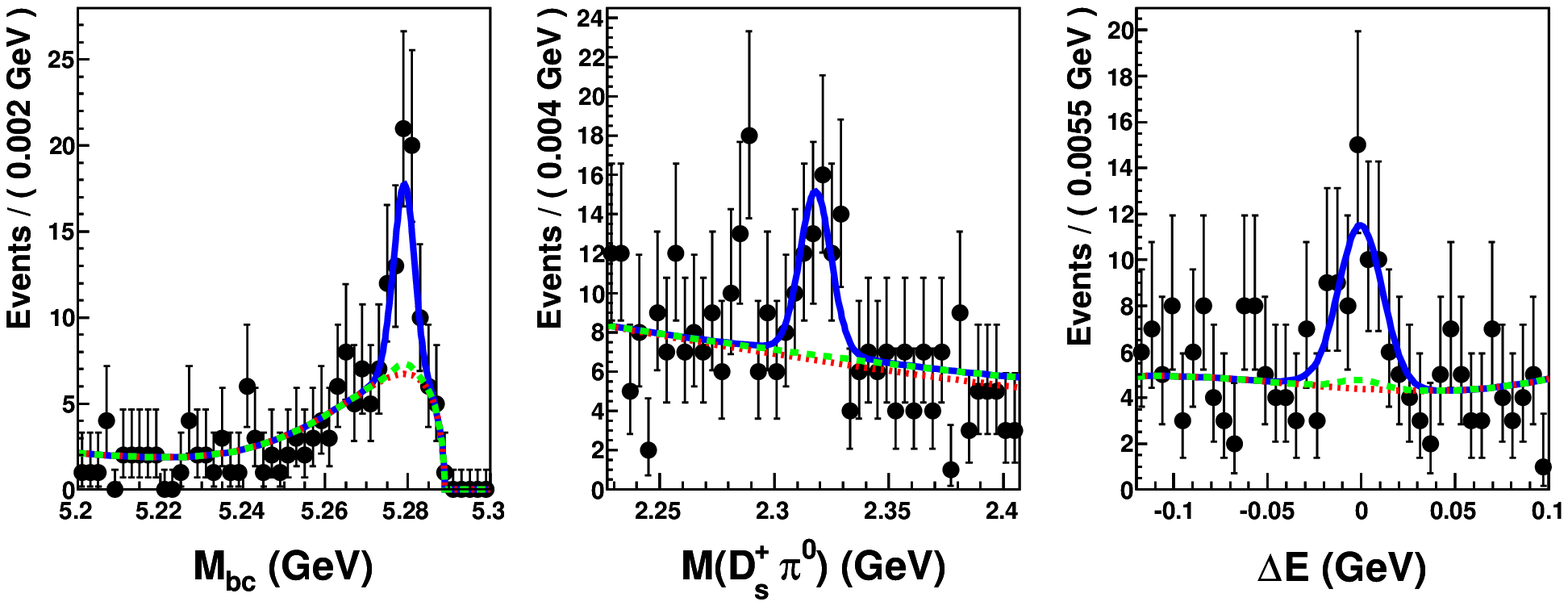}
}
\mbox{
  \includegraphics[height=0.24\textwidth,width=0.47\textwidth]{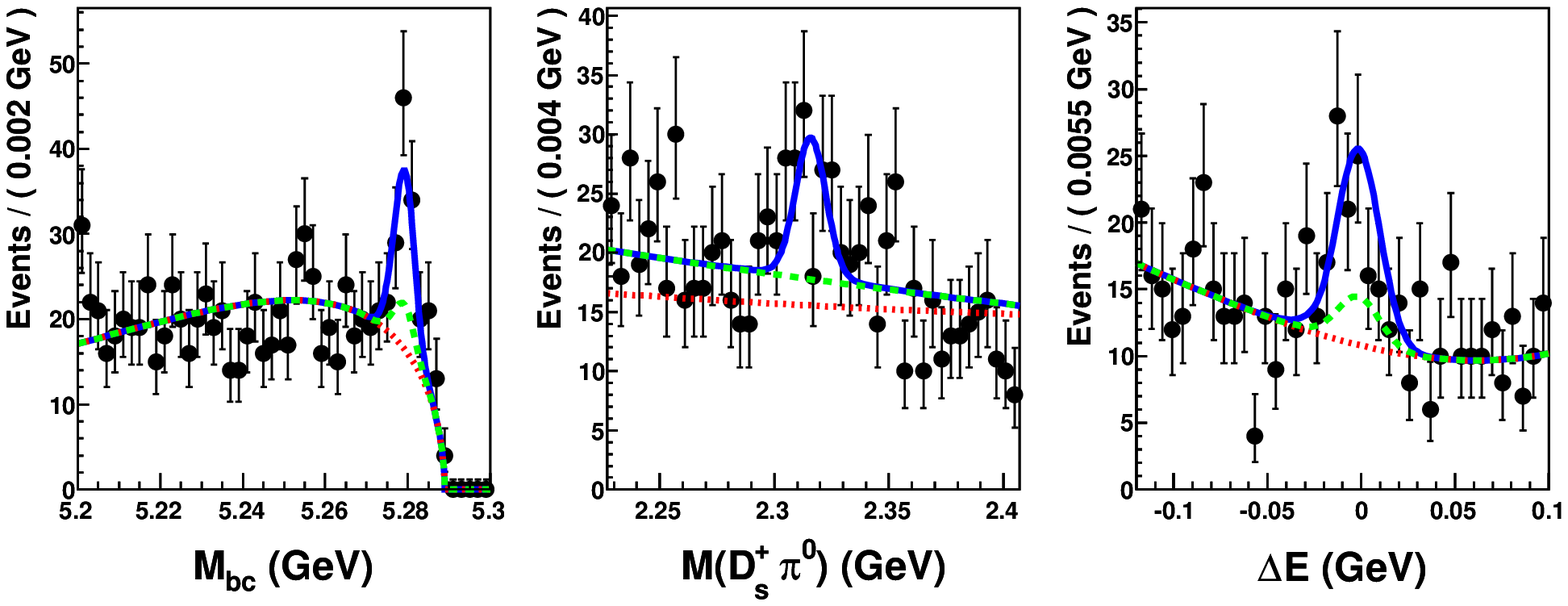}
}
\caption{ {\bf Top:}
The $\Mbc$ (left), $M(\Ds\piz)$ (center) and $\DE$ (right) distributions
for the $\Bpl\rt\Dzbar\Dsz$ candidate events for the $\Dzbar\rt K^+\pim$ subdecay mode,
with the results of the fit superimposed. The events in each distribution are
in the signal regions of the two quantities not being plotted.
{\bf Bottom:} The corresponding distributions for $\Dzbar\rt K^+\pipi\pim$ 
decays. (See text for curves.) 
}

\label{fig:bp-data-fits}
\end{figure}

The top plots of Fig.~\ref{fig:bp-data-fits} show the 
$\Mbc$, $M(\Ds\piz)$ and $\DE$ distributions of the 
$\Bpl\rt\Dzbar\Dsz$,~$\Dsz\rt\Ds\piz$, $\Dzbar\rt K^+\pim$
candidates.  Here, in
addition to the rms widths, we fix the $\Mbc$ and $\DE$ peak positions.
The fit results 
are $38.9^{+9.0}_{-8.2}$ signal events and $12.6^{+22.6}_{-7.7}$ peaking
background events.   An application of the equivalent of Eq.~(\ref{eqn:bf})
to this mode results in the product branching fraction
\begin{eqnarray}
&&{\mathcal B}(\Bpl\rt\Dzbar\Dsz)\times{\mathcal B}(\Dsz\rt\Ds\piz)\\
&=&(7.5 ^{+1.7}_{-1.6} \pm 0.7 \pm 0.3)\times 10^{-4}, 
\nonumber
\end{eqnarray}
which is in good agreement with the PDG average of previous measurements
but with a smaller error. 

The bottom plots of Fig.~\ref{fig:bp-data-fits} show the
$\Mbc$, $M(\Ds\piz)$ and $\DE$ distributions of the 
$\Bpl\rt\Dzbar\Dsz$,~$\Dsz\rt\Ds\piz$, $\Dzbar\rt K^+\pipi\pim$ candidates.
Here again, in
addition to the rms widths, we fix the $\Mbc$ and $\DE$ peak positions.
The fit results 
are $52.4^{+12.5}_{-11.6}$ signal events 
and $99.0^{+12.5}_{-19.9}$ peaking background events.
An application of the equivalent of Eq.~(\ref{eqn:bf})
to this mode results in the product branching fraction
\begin{eqnarray}
&&{\mathcal B}(\Bpl\rt\Dzbar\Dsz)\times{\mathcal B}(\Dsz\rt\Ds\piz)\\
&=&(8.6 ^{+2.1}_{-1.9} \pm 1.1 \pm 0.4)\times 10^{-4}, 
\nonumber
\end{eqnarray}
which is in good agreement with the result for the
$\Dzbar\rt K^+\pim$ mode and the PDG average of previous measurements
and with a comparable error. 

The weighted average of the two measurements is
\begin{eqnarray}
&&{\mathcal B}(\Bpl\rt\Dzbar\Dsz)\times{\mathcal B}(\Dsz\rt\Ds\piz)\\
&=&(8.0 ^{+1.3}_{-1.2} \pm 1.1 \pm 0.4)\times 10^{-4},
\nonumber
\end{eqnarray}
where near-complete correlation of the systematic errors for the two
measurements is taken into account.

As a consistency check, we apply a simultaneous fit to the two modes, where we find a total
signal yield of $91.9^{+15.3}_{-14.6}$ with a statistical significance
of $5.9\sigma$.  The peaking background yield is $148.5 ^{+25.7}_{-24.5}$ events.
The signal yield from the simultaneous fit is consistent with the sum of individual fits,
while the number of peaking background events is marginally higher.  The product branching
fraction obtained using the simultaneous fit is
\begin{eqnarray}
&&{\mathcal B}(\Bpl\rt\Dzbar\Dsz)\times{\mathcal B}(\Dsz\rt\Ds\piz)\\
&=&(8.1  ^{+1.4}_{-1.3} \pm 1.1 \pm 0.3)\times 10^{-4},
\nonumber
\end{eqnarray}
in good agreement with the result from the weighted average 
of results for each mode.

\subsection{3)  Systematic errors}

Systematic errors include the errors on $N_{\bbar}$ and
the $D$ and $\Ds$ secondary branching fractions, MC 
statistics and model dependence, MC-data differences
in particle identification, charged-particle tracking, $\piz$
identification, and the choice of the fitting model.
The error on $N_{\bbar}$ is 1.4\% and the secondary
branching fraction relative errors are the PDG values:
$D^+\rt K^-\pip\pip$ (2.0\%); $D^0\rt K^-\pip$ (1.3\%);
$D^0\rt K^-\pip\pipi$ (2.6\%); $\Ds\rt K^+K^-\pip$ (3.9\%).
The MC model dependence is evaluated by varying 
the $\Ds\rt\phi\pip$ component of $\Ds\rt K^+K^-\pip$
decays between extreme limits and changing the phase-space
distributions for the multibody $D$-meson decay
modes.   
We use various control samples to determine MC-data efficiency differences
that are common to many Belle analyses to evaluate
systematic errors associated with kaon (pion) identification of
1.1\% per track (1.2\% per track), charged particle tracking of 
0.35\% per track, and $\piz$ detection of 4.0\%.

The dependence on the
fitting model is estimated from changes observed by redoing
the fits with each parameter fixed at $\pm 1\sigma$ from its
best-fit value.
The systematic errors from each source, listed in
Table~\ref{tbl:syst},
are summed in quadrature to get the final value.

\begin{table}[htb]
\begin{center}
\caption{\label{tbl:syst}
Summary of relative systematic error sources (in percent).}  
\begin{tabular}{l|c|c|c}\hline  
                         &$\Bz\rt\Dmi\Dsz$         & $\Bpl\rt\Dzbar\Dsz$   &  $\Bpl\rt\Dzbar\Dsz$   \\
                         & $\Dmi\rt K\pi\pi$   &  $\Dzbar\rt K\pi$   & $\Dzbar\rt K 3\pi$    \\
\hline\hline 
$D\&\Ds$ BFs             &           4.4           &        4.1             &   4.7                  \\
\hline\hline 
$N_{\bbar}$               &           1.4           &        1.4             &   1.4                   \\
MC model dep.            &           3.6           &        2.3             &   5.9                   \\
MC stat.                 &           1.2           &        1.0             &   1.4                   \\
Particle ID                &           6.9           &        5.2             &   8.4                   \\
Tracking                 &           2.1           &        1.8             &   2.5                   \\
Fit params.              &           4.4           &        5.8             &   4.7                   \\
$\piz$                   &           4.0           &        4.0             &   4.0                   \\
\hline 
Quad. sum                &          10.2           &        9.4             &  12.4                 \\
\hline 

\hline\hline
\end{tabular}
\end{center}
\end{table}

\section{Search for   $\mathbf  {z^{++}}$ {\boldmath $\rt \Ds\pip$} and $\mathbf  {z^{0}}$ {\boldmath $\rt\Ds\pim$}}

We look for 
$\Zpp\rt\Ds\pip$ and $\Zz\rt \Ds\pim$ signals in the $B^+\rt D^- \Ds\pip$ and $B^0\rt\Dzbar\Ds\pim$
decay channels by applying the selection criteria discussed above with the
replacement of  the selected $\pi^0$ with a $\pi^+$ (for $\Zpp$) or $\pim$ (for $\Zz$).
Here, for events with multiple $\bar{D}$ and/or $\Ds$ track combinations, we select those with 
a measured invariant mass closest to the corresponding PDG values. 
For $\Zpp$ signal MC, the number of remaining events with multiple candidates is
11.2\% over the full three-dimensional range of the likelihood fit; for $\Zz$, fewer than $0.1\%$ 
of the remaining events have multiple candidates.

\subsection{1)  Peaking backgrounds from generic MC samples}
\begin{figure}[htb]
\mbox{
  \includegraphics[height=0.24\textwidth,width=0.47\textwidth]{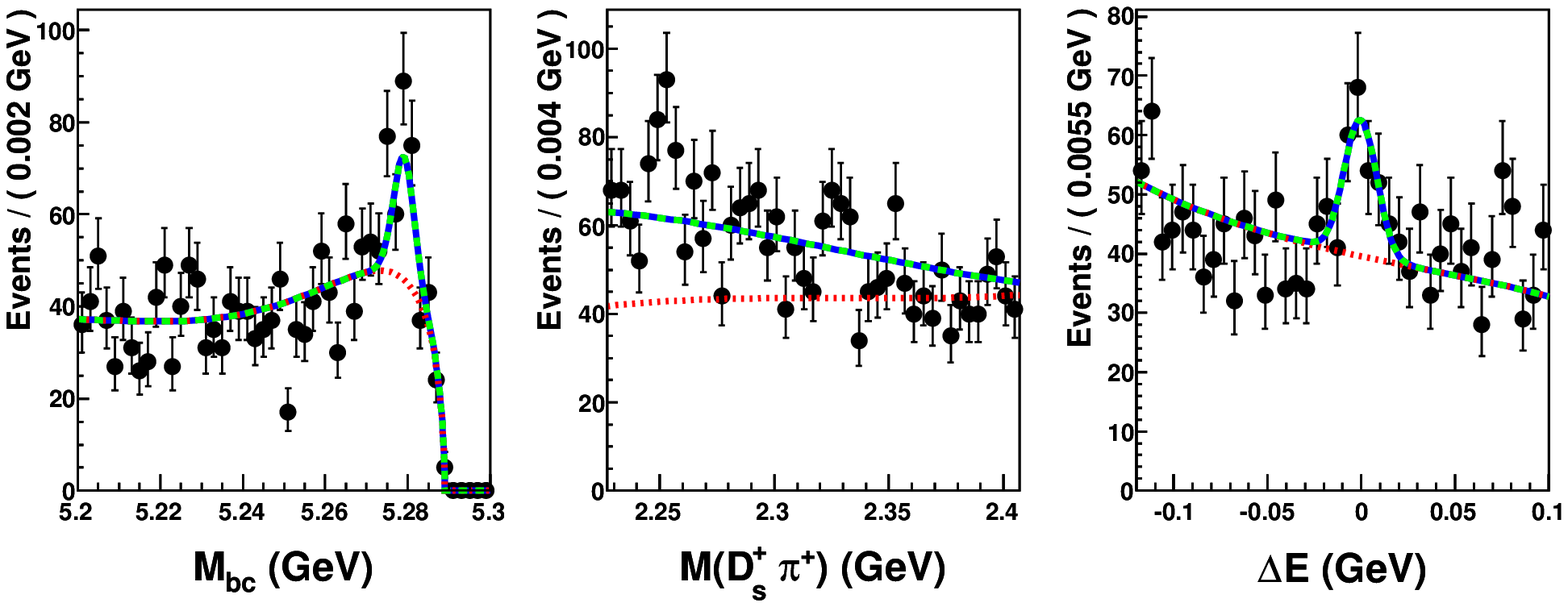}
}
\mbox{
  \includegraphics[height=0.24\textwidth,width=0.47\textwidth]{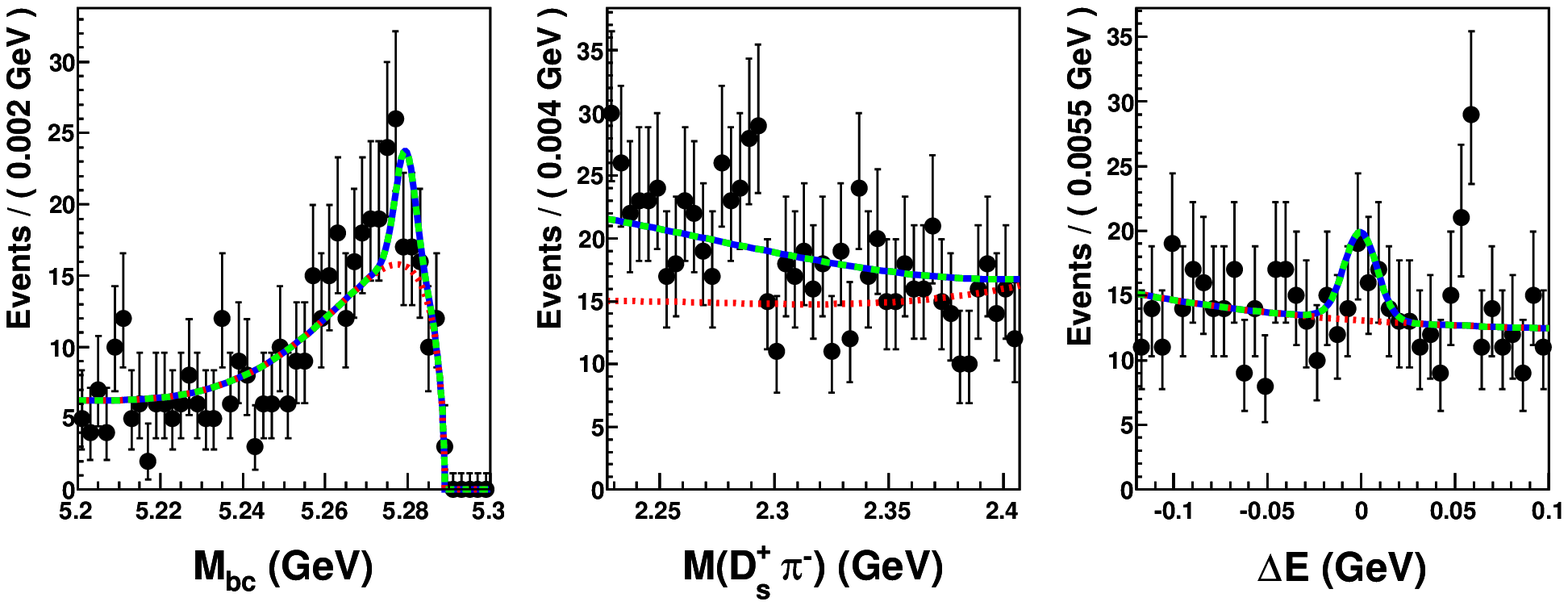}
}
\mbox{
  \includegraphics[height=0.24\textwidth,width=0.47\textwidth]{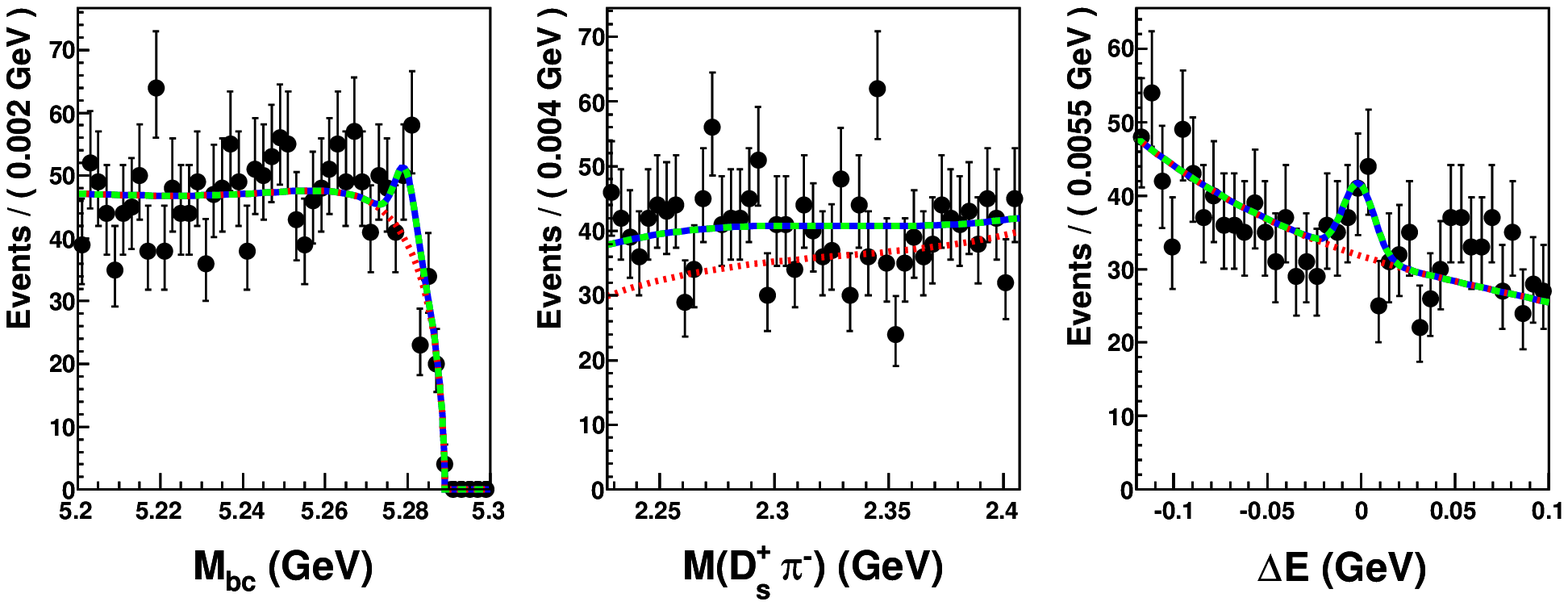}
}

\caption{ 
The $\Mbc$ (left), $M(\Ds\pi)$ (center) and $\DE$ (right) distributions 
for generic-MC events  that pass the $\Dmi\Ds\pip$ (top), $\Dzbar\Ds\pim$,
$\Dzbar\rt K^+\pim$ (middle) and $\Dzbar\Ds\pim$, $\Dzbar\rt K^+\pipi\pim$
(bottom) channels.  The curves are the results of fits described in the text.}

\label{fig:dbl-gen-mc-fits}
\end{figure}

We check for possible peaking backgrounds leaking into the signal using
a sample of simulated generic $B$-meson decay events (with no
$\Zpp$ or $\Zz$ signals) with a luminosity that corresponds
to 3 times the number of $B$ decays in the data.  
The top plots of Fig.~\ref{fig:dbl-gen-mc-fits}
show the results of applying the three-dimensional fit to 
selected $\Dmi\Ds\pip$ MC events.
Here, the signal yield is zero with a positive error of 7.1 events.
The peaking background yield is $544\pm 41$ events.
The middle (bottom) plots of Fig.~\ref{fig:dbl-gen-mc-fits} show the results of the 
three-dimensional fits to the generic MC for the $\Dzbar\rt K^+\pim$ 
($\Dzbar\rt K^+\pipi\pim$) channel in the selected $B\rt\Zz\Dzbar$ samples.
No background processes are found that produce a spurious signal; 
the signal yields are also zero for both $\Dzbar$ modes with positive errors
of 2.1 and 9.9 events for the $K^+\pim$ and $K^+\pipi\pim$ modes, respectively.
The $\Mbc$-$\DE$ peaking background yields for these modes are $169\pm 22$ and
$229^{+32}_{-31}$ events, respectively.

\subsection{2)  Mass-dependent efficiency}

Since the $\Zpp$ and $\Zz$ are hypothesized 
to be isospin partners of the $\Dsz$,
their masses are expected to lie somewhere within a $\pm 10$~MeV mass region of
$m_{\Dsz}=2317.8\pm 0.6$~MeV.  In order to be certain that we cover all
reasonably plausible mass values, we scan for $\Zpp$ and $\Zz$ signals in 13 
adjacent mass bins, each 5~MeV wide, covering a $\pm 32.5$~MeV interval centered on $2317.8$~MeV.

To account for a possible mass dependence of the detection efficiency, we generate
$\Zpp$ and $\Zz$ signal MC events with $\mathbf{z}$ masses 
in the full range of the scan.
The efficiencies, determined from fits to the selected events from each MC sample,
are independent of mass to within the
$\sim$2.5\% MC statistical errors.  For the $\Zpp$ search, the average efficiency is
($8.3\pm 0.1$)\%.  For the $\Zz$ search, the average efficiency is ($9.2\pm 0.1$)\% 
for the $\Dzbar\rt K^+\pim$ mode and ($4.1\pm 0.1$)\% for $\Dzbar\rt K^+\pipi\pim$.

\subsection{3)  Fits to the {\boldmath $ M(\Ds\pi^{+,-})$}  spectra}

\begin{figure}[htb]
\mbox{
  \includegraphics[height=0.24\textwidth,width=0.47\textwidth]{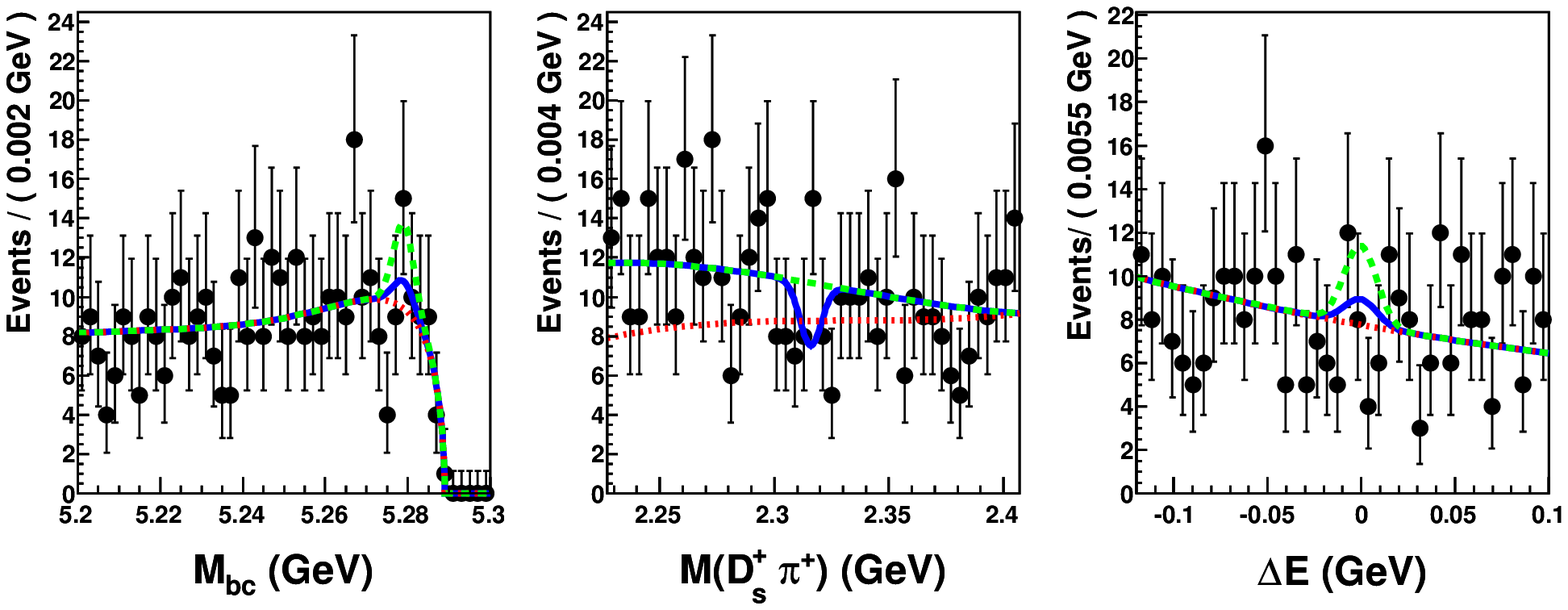}
}
\mbox{
  \includegraphics[height=0.24\textwidth,width=0.47\textwidth]{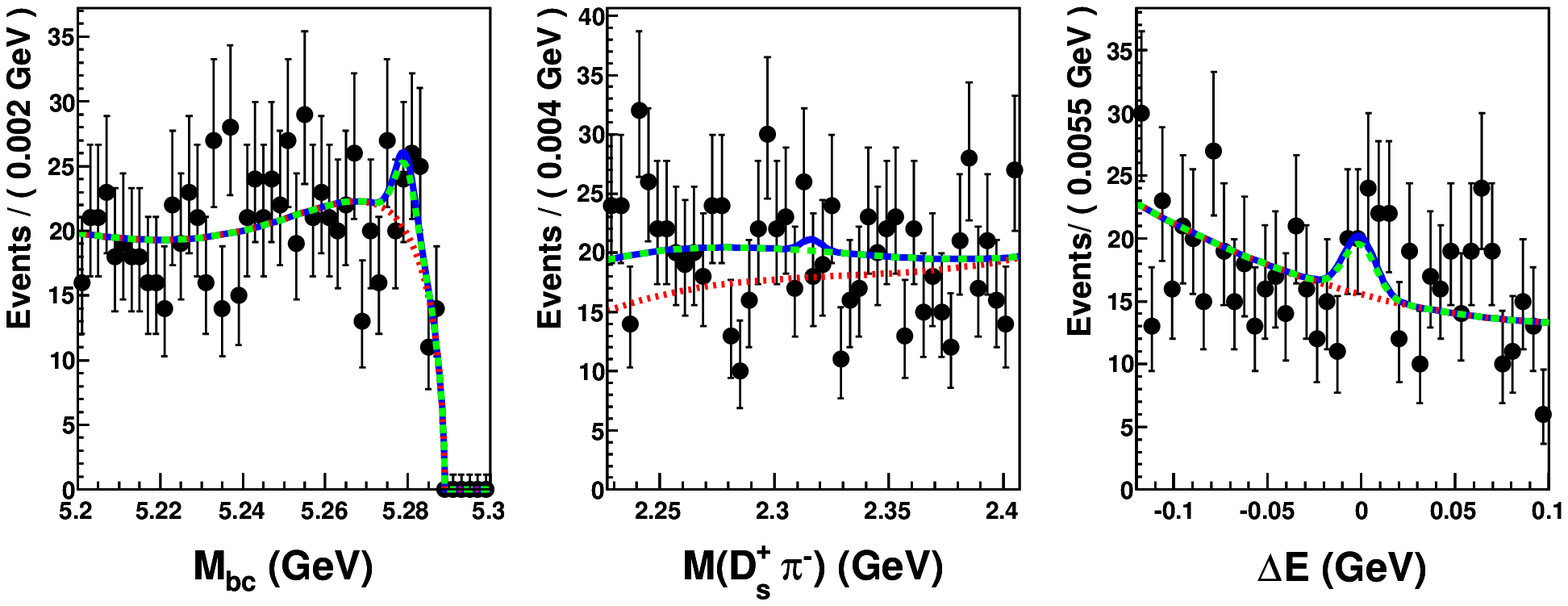}
}
\caption{ 
The $\Mbc$ (left), $M(\Ds\pi)$ (center) and $\DE$ (right) distributions for selected
$B^{+}\rt \Dmi\Ds\pip $ (top) and $B^0\rt \Dzbar\Ds\pim$ (bottom) event candidates
for the fit with the signal peak mass restricted to a 5~MeV region centered at
$M(\Ds\pi)=2317.8$~MeV. In the lower plots, the $\Dzbar\rt K^+\pim$ and $K^+\pipi\pim$
samples are combined. (See text for curves.) 
}
\label{fig:scan}
\end{figure}

We apply a sequence of 13 three-dimensional fits to the data using a Gaussian signal
function with width fixed at the MC-determined $D_s^{+} \pi^{\pm}$ mass 
resolution ($\sigma$=4.6~MeV) to represent the $\Zpp$ ($\Zz$) 
with a peak mass restricted to 5~MeV-wide windows
%
covering a total mass range of $\pm 32.5$~MeV about $m_{\Dsz}=2317.8$~MeV.
The results of these fits for the
$\Zpp\rt\Ds\pip$ and $\Zz\rt\Ds\pim$ searches are summarized in Table~\ref{tbl:scan}. 
As examples, we show the fit results for the mass bin centered at $M(\Ds\pi)=2317.8$~MeV
for the $\Zpp$ ($\Zz$) search in the top (bottom) plots of Fig.~\ref{fig:scan}. 
None of the fits returns  a positive $\Zpp$ or $\Zz$ signal with a statistical significance
of more than $1.3\sigma$.  The determination of the Bayesian 90\% 
credibility level upper limits~\footnote{
Common convention has used the frequentist label ``confidence level''
for this criterion.}
on the event yields and
product branching fractions is described below.

\begin{table}[htb]
\begin{center}
\caption{\label{tbl:scan}
Product branching fraction upper limits ${\mathcal B}^{\rm UL}_{i}$
for ${\mathcal B}(B^+~(B^0 )~\rt\Dmi~(\Dzbar)~z_i)\times {\mathcal B}(z_i~\rt\Ds\pi~)$
($z_{1}=\Zpp$ and $z_{2}=\Zz$), for $z_{i}$ masses between 2285.3~MeV and 2350.3~MeV. 
Here $\Delta M=M_{\rm ctr}-m_{\Dsz}$, where $M_{\rm ctr}$ is the center of the 5~MeV mass 
window allowed for the fit, and $N^{\rm UL}_{i}$  is the upper limit
including systematic errors.}    
\begin{tabular}{l|c|c|c|c|c|c}\hline  
$\Delta M$&$N^{\rm fit}_{++}$&~$N^{\rm UL}_{++}$~&${\mathcal B}^{\rm UL}_{++}$&$N^{\rm fit}_{0}$&~$N^{\rm UL}_{0}$~&${\mathcal B}^{\rm UL}_{0}$\\
      MeV &           &              &    ($10^{-4}$)   &             &              &    ($10^{-4}$)   \\
\hline\hline 
   $-30$& $4.0^{+5.9}_{-9.0}$  & $16.3$ & 0.52 & $-13.7\pm 6.2$    &~$10.5$~&~0.34~\\ 
   $-25$ & $4.1^{+5.9}_{-5.9}$  & $16.3$ & 0.52 & $5.5^{+7.9}_{-15.6}$~& $21.2$ & 0.69  \\ 
   $-20$ & $-8.3^{+5.3}_{-4.2}$ &  $9.8$ & 0.32 & $5.8^{+8.0}_{-8.2}$  & $21.5$ & 0.69  \\
   $-15$ & $-10.3^{+4.0}_{-3.1}$&  $8.0$ & 0.25 & $2.7\pm 8.3$      & $20.1$ & 0.65 \\
   $-10$ & $-10.2\pm 3.5$    &  $7.9$ & 0.25 & $4.0^{+7.9}_{-8.4}$  & $20.4$ & 0.66 \\
   $ -5$ & $-8.8\pm 3.2$     &  $8.1$ & 0.25 & $4.1\pm 7.4$      & $20.4$ & 0.66 \\
   $~~0$ & $-9.3\pm 3.0$     &  $8.4$ & 0.27 & $3.1^{+7.8}_{-7.9}$  & $19.8$ & 0.64 \\
   $~~5$ & $-9.3^{+4.5}_{-3.0}$ &  $8.5$ & 0.28 & $-1.7^{+10.3}_{-6.1}$ & $16.0$ & 0.52 \\
   $~10$ & $4.6^{+5.6}_{-10.8}$ & $16.2$ & 0.51 & $-5.4^{+7.6}_{-5.3}$  & $13.4$ & 0.44 \\
   $~15$ &~$6.4\pm 5.0$       & $17.8$ & 0.57 & $-5.4^{+6.7}_{-5.3} $ & $13.3$ & 0.43 \\
   $~20$ & $6.0^{+5.9}_{-5.1}$  & $17.6$ & 0.56 & $-3.3^{+11.5}_{-5.6}$ & $14.3$ & 0.47 \\
   $~25$ & $3.0^{+6.9}_{-5.9}$  & $15.8$ & 0.50 & $5.7^{+7.2}_{-6.9}$  & $20.6$ & 0.67 \\
   $~30$ & $3.4^{+5.7}_{-5.6}$  & $15.8$ & 0.50 & $5.6^{+7.0}_{-5.1}$  & $20.0$ & 0.65 \\
\hline\hline 
\end{tabular}
\end{center}
\end{table}

\subsection{4)  Systematic errors for $\Zpp$ and $\Zz$ searches}

Systematic errors are evaluated using the same methods that are used for the
$\Dsz$ branching fraction measurement described above, with the $\piz$-associated
error replaced by the error on the additional charged pion. For this, the nominal
0.35\% tracking error is assigned to $p>200$~MeV tracks.  However, 5\% of the
relevant pions for the $\Zz$ have $p<200$~MeV with an associated error of 5\%. Here, a weighted
average is used and the total tracking uncertainty increases to 3.8\%.  For the
systematic error associated with multiple candidates, we perform a multiple-candidate-free
$\Zpp$ scan where we use the smallest $\DE$ to select the best candidate and
a two-dimensional fit [$\Mbc$ and $M(\Ds\pip)$] to measure signal yields. From
the differences between the results of the two methods, we determine a systematic
error from this source of 2.2\%. For other sources  of error, we use the results
listed in Table~\ref{tbl:syst}.   The resulting errors
are 11.4\% for the $\Zpp$ search and 16.6\% for the $\Zz$ search.

\subsection{5)  Upper limit determination}

We use a Bayesian method to convert the fitted results to upper limits on
the total number of signal events.  To account for the systematic uncertainties, 
the likelihood distributions from the $\Zpp$ ($\Zz$), fits are convolved with 
a Gaussian with  $\sigma_{\rm syst}=0.114~(0.166) \times N^{\rm UL}_{\rm stat}$, 
where $N^{\rm UL}_{\rm stat}$ is determined from
\begin{equation}
\frac{\int_0^{N^{\rm UL}_{\rm stat}} {\mathcal L}(n_{\rm sig}) dn_{\rm sig}}
{\int_0^{+\infty} {\mathcal L}(n_{\rm sig}) dn_{\rm sig}} =0.9.
\label{eqn:like-stat}
\end{equation}
The Gaussian width is $\sigma_{\rm syst} =1.1~(3.1)$~events for the 2317.8~MeV mass bin
of the $\Zpp$ ($\Zz$) scan; the widths for the other mass bins are similar.
The corresponding upper limits, $N^{\rm UL}$, are determined from the relation
\begin{equation}
\frac{\int_{0}^{N^{\rm UL}}{\mathcal L}(n_{\rm sig})\bigotimes {\mathcal G}(n_{\rm sig})dn_{\rm sig}}
{\int_{0}^{+\infty} {\mathcal L}(n_{\rm sig})\bigotimes {\mathcal G}(n_{\rm sig})dn_{\rm sig}} =0.9,
\label{eqn:like-conv}
\end{equation}
and in all cases differ from $N^{\rm UL}_{\rm stat}$ by less than one event. 
The resulting values of $N^{\rm UL}$ are listed in Table~\ref{tbl:scan}.

For the $\Zpp$ search, we determine
upper limits on the product branching fractions 
${\mathcal B}^{\rm UL}_{++}\equiv{\mathcal B}(B\rt \Dmi\Zpp)\times{\mathcal B}(\Zpp\rt\Ds\pip)$ 
from the relation

\begin{equation}
{\mathcal B}^{\rm UL}_{++}   
= \displaystyle{}\frac{N^{\rm UL}_{++}}
{N_{\bbar} {\mathcal B}_{\Ds}{\mathcal B}_{\Dmi}\eta_{++}}, 
\end{equation}
where the notation follows that of Eq.~(\ref{eqn:bf}) 
and $\eta_{++}$ is the MC-determined efficiency.
For the $\Zz$ search, where there is no evidence for the signal either,
we use the same relation with ${\mathcal B}_{\Dmi}\eta_{++}$ replaced by
${\mathcal B}_{K\pi}\eta_{K\pi} +
{\mathcal B}_{K3\pi}\eta_{K3\pi}$,
where $\eta_{K\pi}$ ($\eta_{K3\pi}$) is the efficiency for the
$\Dzbar\rt K^+\pim$ ($K^+\pipi\pim$) mode.  The resulting  product
branching fraction upper limits, listed in Table~\ref{tbl:scan}, are
all more than an order of magnitude lower than the measured values for
the $\bar{D}\Dsz$ final states.  This is in strong contradiction to
expectations for the hypothesis that the $\Dsz$ is a member of an isospin triplet~\cite{hayashigaki}.

\section{Summary}
We reported measurements of the product branching fractions 
${\mathcal B}(\Bpl\rt\Dzbar\Dsz)\times {\mathcal B}(\Dsz\rt\Ds\piz)
=(8.0^{+1.3}_{-1.2} \pm 1.1 \pm 0.4)\times 10^{-4}$
and 
${\mathcal B}(\Bz\rt\Dmi\Dsz)\times{\mathcal B}(\Dsz\rt\Ds\piz)
=(10.2^{+1.3}_{-1.2} \pm 1.0 \pm 0.4)\times 10^{-4}$.  Here, the 
first errors are statistical, the second are systematic and
the third are from $D$ and $\Ds$ branching fractions.  These
values agree with the existing PDG world average values~\cite{pdg},
significantly improve upon their precision, and 
supersede those of Ref.~\cite{belle_bdsz}.
In addition,  we reported negative results on a search for
hypothesized doubly charged and neutral isospin partners of the $\Dsz$
and provided upper limits on the product branching fractions
that are more than an order of magnitude smaller than the
theoretical predictions of Hayashigaki and
Terasaki~\cite{hayashigaki}.

\section{Acknowledgments}

We thank the KEKB group for the excellent operation of the
accelerator; the KEK cryogenics group for the efficient
operation of the solenoid; and the KEK computer group,
the National Institute of Informatics, and the 
PNNL/EMSL computing group for valuable computing
and SINET4 network support.  We acknowledge support from
the Ministry of Education, Culture, Sports, Science, and
Technology (MEXT) of Japan, the Japan Society for the 
Promotion of Science (JSPS), and the Tau-Lepton Physics 
Research Center of Nagoya University; 
the Australian Research Council and the Australian 
Department of Industry, Innovation, Science and Research;
Austrian Science Fund under Grant No.~P 22742-N16 and P 26794-N20;
the National Natural Science Foundation of China under Contracts 
No.~10575109, No.~10775142, No.~10875115, No.~11175187, and  No.~11475187; 
the Ministry of Education, Youth and Sports of the Czech
Republic under Contract No.~LG14034;
the Carl Zeiss Foundation, the Deutsche Forschungsgemeinschaft
and the VolkswagenStiftung;
the Department of Science and Technology of India; 
the Istituto Nazionale di Fisica Nucleare of Italy; 
National Research Foundation (NRF) of Korea Grants
No.~2011-0029457, No.~2012-0008143, No.~2012R1A1A2008330, 
No.~2013R1A1A3007772, No.~2014R1A2A2A01005286, No.~2014R1A2A2A01002734, 
No.~2014R1A1A2006456;
the Basic Research Lab program under NRF Grant No.~KRF-2011-0020333, 
No.~KRF-2011-0021196, Center for Korean J-PARC Users, No.~NRF-2013K1A3A7A06056592; 
the Brain Korea 21-Plus program and the Global Science Experimental Data 
Hub Center of the Korea Institute of Science and Technology Information;
the Institute of Basic Science (Korea) Project Code
IBS-DR016-D1;
the Polish Ministry of Science and Higher Education and 
the National Science Center;
the Ministry of Education and Science of the Russian Federation and
the Russian Foundation for Basic Research;
the Slovenian Research Agency;
the Basque Foundation for Science (IKERBASQUE) and 
the Euskal Herriko Unibertsitatea (UPV/EHU) under program UFI 11/55 (Spain);
the Swiss National Science Foundation; the National Science Council
and the Ministry of Education of Taiwan; and the U.S.\
Department of Energy and the National Science Foundation.
This work is supported by a Grant-in-Aid from MEXT for 
Science Research in a Priority Area (``New Development of 
Flavor Physics'') and from JSPS for Creative Scientific 
Research (``Evolution of Tau-lepton Physics'').


\begin{thebibliography}{99}
\bibitem{conj}
In this report the inclusion of charge-conjugate states
is always implied.
\end{thebibliography}
\end{document}